\documentclass[twocolumn,trackchanges]{aastex62}

\usepackage{epsfig,graphicx,amssymb,epstopdf,mathrsfs,float,url}
\newcommand{\egam}{\varepsilon_\gamma}

\newcommand{\fermi}{{\em Fermi}}

\newcommand{\ice}{IceCube}

\newcommand{\antares}{ANTARES}

\newcommand{\txsblazar}{TXS~0506+056}

\newcommand{\nugamma}{\mbox{$\nu$+$\gamma$}}

\newcommand{\lamc}{\mbox{$\lambda_{\rm C}$}}
\newcommand{\lamd}{\mbox{$\lambda_{\rm D}$}}
\newcommand{\ninj}{\mbox{$n_{\rm inj}$}}
\newcommand{\ngexp}{\mbox{$\langle n_\gamma \rangle$}}

\def\arcdeg{\hbox{$^\circ$}}
\def\arcmin{\hbox{$^\prime$}}

\def\simlt{\mathrel{\hbox{\rlap{\hbox{\lower4pt\hbox{$\sim$}}}\hbox{$<$}}}}
\def\simgt{\mathrel{\hbox{\rlap{\hbox{\lower4pt\hbox{$\sim$}}}\hbox{$>$}}}}

\newcommand{\gray}{\mbox{$\gamma$-ray}}
\newcommand{\grays}{\mbox{$\gamma$-rays}}

\newcommand{\peryear}{{{\rm yr}$^{-1}$}}

\begin{document}

\title{A Search for Cosmic Neutrino and Gamma-Ray Emitting Transients
  in 7.3 Years of ANTARES and \fermi\ LAT Data}

\correspondingauthor{Colin Turley}
\email{cft114@psu.edu}


\author{H.~A. Ayala Solares}
\affiliation{Department of Physics, Pennsylvania State University,
  University Park, PA 16802, USA}
\affiliation{Center for Particle \& Gravitational Astrophysics,
  Institute for Gravitation and the Cosmos, PennsylvaniaState
  University, University Park, PA 16802, USA}

\author{D.~F. Cowen}
\affiliation{Department of Physics, Pennsylvania State University,
  University Park, PA 16802, USA}
\affiliation{Department of Astronomy \& Astrophysics, Pennsylvania
  State University, University Park, PA 16802, USA}
\affiliation{Center for Particle \& Gravitational Astrophysics,
  Institute for Gravitation and the Cosmos, PennsylvaniaState
  University, University Park, PA 16802, USA}

\author{J.~J. DeLaunay}
\affiliation{Department of Physics, Pennsylvania State University,
  University Park, PA 16802, USA}
\affiliation{Center for Particle \& Gravitational Astrophysics,
  Institute for Gravitation and the Cosmos, PennsylvaniaState
  University, University Park, PA 16802, USA}

\author{D.~B. Fox}
\affiliation{Department of Astronomy \& Astrophysics, Pennsylvania
  State University, University Park, PA 16802, USA}
\affiliation{Center for Particle \& Gravitational Astrophysics,
  Institute for Gravitation and the Cosmos, PennsylvaniaState
  University, University Park, PA 16802, USA}
\affiliation{Center for Theoretical \& Observational Cosmology,
  Institute for Gravitation and the Cosmos, Pennsylvania State
  University, University Park, PA 16802, USA}

\author{A. Keivani}
\affiliation{Department of Physics, Pennsylvania State University,
  University Park, PA 16802, USA}
\affiliation{Center for Particle \& Gravitational Astrophysics,
  Institute for Gravitation and the Cosmos, PennsylvaniaState
  University, University Park, PA 16802, USA}

\author{ M. Mostaf\'{a}}
\affiliation{Department of Physics, Pennsylvania State University,
  University Park, PA 16802, USA}
\affiliation{Department of Astronomy \& Astrophysics, Pennsylvania
  State University, University Park, PA 16802, USA}
\affiliation{Center for Particle \& Gravitational Astrophysics,
  Institute for Gravitation and the Cosmos, PennsylvaniaState
  University, University Park, PA 16802, USA}

\author{K. Murase}
\affiliation{Department of Physics, Pennsylvania State University,
  University Park, PA 16802, USA}
\affiliation{Department of Astronomy \& Astrophysics, Pennsylvania
  State University, University Park, PA 16802, USA}
\affiliation{Center for Particle \& Gravitational Astrophysics,
  Institute for Gravitation and the Cosmos, PennsylvaniaState
  University, University Park, PA 16802, USA}

\author{C.~F. Turley}
\affiliation{Department of Physics, Pennsylvania State University,
  University Park, PA 16802, USA}
\affiliation{Center for Particle \& Gravitational Astrophysics,
  Institute for Gravitation and the Cosmos, PennsylvaniaState
  University, University Park, PA 16802, USA}

\collaboration{AMON}

\author{A. Albert}
\affiliation{Universit\'e de Strasbourg, CNRS, IPHC UMR 7178, F-67000
  Strasbourg, France}

\author{M.~Andr\'e}
\affiliation{Technical University of Catalonia, Laboratory of Applied
  Bioacoustics, Rambla Exposici\'o, 08800 Vilanova i la Geltr\'u,
  Barcelona, Spain}

\author{M.~Anghinolfi}
\affiliation{INFN - Sezione di Genova, Via Dodecaneso 33, 16146
  Genova, Italy}

\author{G.~Anton}
\affiliation{Friedrich-Alexander-Universit\"at Erlangen-N\"urnberg,
  Erlangen Centre for Astroparticle Physics, Erwin-Rommel-Str. 1,
  91058 Erlangen, Germany}

\author{M.~Ardid}
\affiliation{Institut d'Investigaci\'o per a la Gesti\'o Integrada de
  les Zones Costaneres (IGIC) - Universitat Polit\`ecnica de
  Val\`encia. C/ Paranimf 1, 46730 Gandia, Spain}

\author{J.-J.~Aubert}
\affiliation{Aix Marseille Univ, CNRS/IN2P3, CPPM, Marseille, France}

\author{J.~Aublin}
\affiliation{APC, Univ Paris Diderot, CNRS/IN2P3, CEA/Irfu, Obs de
  Paris, Sorbonne Paris Cit\'e, France}

\author{B.~Baret}
\affiliation{APC, Univ Paris Diderot, CNRS/IN2P3, CEA/Irfu, Obs de
  Paris, Sorbonne Paris Cit\'e, France}

\author{J.~Barrios-Mart\'{\i}}
\affiliation{IFIC - Instituto de F\'isica Corpuscular (CSIC -
  Universitat de Val\`encia) c/ Catedr\'atico Jos\'e Beltr\'an, 2
  E-46980 Paterna, Valencia, Spain}

\author{S.~Basa}
\affiliation{LAM - Laboratoire d'Astrophysique de Marseille, P\^ole de
  l'\'Etoile Site de Ch\^ateau-Gombert, rue Fr\'ed\'eric Joliot-Curie
  38, 13388 Marseille Cedex 13, France}

% ...

\collaboration{ANTARES Collaboration}


\begin{abstract}

  We analyze 7.3~years of \antares\ high-energy neutrino and
  \fermi\ LAT \gray\ data in search of cosmic neutrino + \gray\
  (\nugamma) transient sources or source populations. 
  Our analysis has the potential to detect either individual
  \nugamma\ transient sources (durations $\delta t \simlt 1000$\,s),
  if they exhibit sufficient \gray\ or neutrino multiplicity, or a
  statistical excess of \nugamma\ transients of individually lower 
  multiplicities. 
  Individual high \gray-multiplicity events could be produced,
  for example, by a single \antares\ neutrino in coincidence
  with a LAT-detected \gray\ burst.
  Treating \antares\ track and cascade event types separately, we
  establish detection thresholds by Monte Carlo scrambling of the
  neutrino data, and determine our analysis sensitivity by signal
  injection against these scrambled datasets. We find our analysis
  is sensitive to \nugamma\ transient populations responsible for
  $>$5\% of the observed gamma-coincident neutrinos in the track
  data at 90\% confidence.
  Applying our analysis to the unscrambled data reveals no individual
  \nugamma\ events of high significance; two \antares\ track +
  \fermi\ \gray\ events are identified that exceed a once per decade
  false alarm rate threshold ($p=17\%$). No evidence for subthreshold
  \nugamma\ source populations is found among the track ($p=39\%$) or
  cascade ($p=60\%$) events.
  Exploring a possible correlation of high-energy neutrino directions
  with \fermi\ \gray\ sky brightness identified in previous
  work yields no added support for this correlation.
  While \txsblazar, a blazar and variable (non-transient)
  \fermi\ \gray\ source, has recently been identified as the first
  source of high-energy neutrinos, the challenges in reconciling
  observations of the \fermi\ \gray\ sky, the \ice\ high-energy cosmic
  neutrinos, and ultra-high energy cosmic rays using only blazars
  suggest a significant contribution by other source populations.
  Searches for transient sources of high-energy neutrinos
  thus remain interesting, with the potential for either
  neutrino clustering or multimessenger coincidence searches to lead
  to discovery of the first \nugamma\ transients. 
   
\end{abstract}



\section{Introduction}
\label{sec:intro}

The \antares\ telescope \citep{antdesg11} is a deep-sea Cherenkov
neutrino detector, located 40~km off shore from Toulon, France, in the
Mediterranean Sea. The detector comprises a three-dimensional array of
885 optical modules, each one housing a \mbox{10\,in} photomultiplier
tube, and distributed over 12 vertical strings anchored in the sea
floor at a depth of about 2400\,m. The detection of light from
up-going charged particles is optimized with the photomultipliers
facing 45$^\circ$ downward.  Completed in May~2008, the telescope aims
primarily at the detection of neutrino-induced muons that cause the
emission of Cherenkov light in the detector (\textit{track-like}
events). Charged current interactions induced by electron neutrinos
(and, possibly, by tau neutrinos of cosmic origin) or neutral current
interactions of all neutrino flavors can be reconstructed as
\textit{cascade-like} events \citep{Albert:2017hyy}.

Due to its location, the \antares\ detector mainly observes the
Southern sky ($2\pi$\,sr at any time).  Events arising from sky
positions in the declination band $-90^\circ \le \delta \le -48^\circ
$ are always visible as upgoing.  Neutrino-induced events in the
declination band $-48^\circ \le \delta \le +48^\circ $ are visible as
upgoing with a fraction of time decreasing from 100\% down to 0\%.
While \antares\ has a substantially smaller volume than IceCube, the
use of sea water as detection medium (rather than ice) provides better
pointing resolution for individual events, especially those of cascade
type, and its geographic location enables reduced-background studies
of the Southern hemisphere including the Galactic center region. On
the other hand, natural light emission in the water leads to higher
background levels \citep{antlight04}.

Chief scientific results from \antares\ include searches for neutrino
sources using track- and cascade-like events in data collected between
2007 and 2015 \citep{Albert:2017ohr}; dedicated studies along the
Galactic Plane \citep{Albert:2017oba}, also in collaboration with the
IceCube telescope \citep{ic-ant2018}; searches for an excess of
high-energy cosmic neutrinos over the background of atmospheric events
\citep{antcosmic9y}. No cosmic neutrinos have been positively
identified in the \antares\ data. Despite this, by integrating the
cosmic neutrino spectrum from \citet{icecubeicrc17} over the
\antares\ effective area \citep{Albert:2017ohr}, we estimate an
expected 6.8 neutrinos of cosmic origin are detected each year,
though all but the most energetic will be indistinguishable from the
atmospheric background. Among all the possible astrophysical
sources, transient sources increase the observation possibilities
thanks to the suppression of atmospheric background in a well-defined
space-time window.  For this reason, the Collaboration is involved in
a broad multimessenger program to exploit the connection between
neutrinos and other cosmic messengers, including: follow-up analyses
associated with gravitational wave events
\citep{ANTARES:2017bia,Albert:2018jnn}; coincidence searches against
electromagnetic observations from radio
\citep{Croft:2016lhf,Albert:2018euo} and visible
\citep{Adrian-Martinez:2015nin} to X- and
\grays\ \citep{Ageron:2011pe}; blazar flare episodes
\citep{Adrian-Martinez:2015wis}; and the neutrino source
\txsblazar\ \citep{anttxs18}.  To date, there have been no
high-confidence counterparts identified for any \antares\ neutrino
event.

In parallel, members of the Astrophysical Multimessenger Observatory
Network (AMON\footnote{AMON website: \url{http://www.amon.psu.edu/}};
\citealt{amondesg,amonrt}) have been exploring the possibility of
neutrino + \gray\ (\nugamma) source identification via coincidence
analysis, publishing analyses of \fermi\ Large Area Telescope (LAT;
\citealt{latdesg}) and public \ice\ \mbox{40-string}
\citep{keivani2015} and \mbox{59-string} \citep{turleyfermi18}
data. Although no high-confidence \nugamma\ transients, nor evidence
of subthreshold \nugamma\ source populations, were identified in these
works, the latter revealed mild evidence for correlation between
\ice\ neutrino positions and the \fermi\ \gray\ sky.

Within the last year, a coincidence between the neutrino
\mbox{IceCube-170922A} \citep{170922gcn} and the flaring blazar
\txsblazar\ \citep{1709blazaratel} led to multimessenger
\citep{blazarnu} and time-dependent neutrino clustering
\citep{txscluster18} analyses suggesting this BL~Lac-type object as
the first known source of high-energy neutrinos and the first
identified extragalactic cosmic ray accelerator. Further blazar source
identifications can certainly be anticipated; however, the absence of
point source excesses in the \antares\ \citep{Albert:2017ohr} and
\ice\ \citep{icpoint2017,ic-ant2018} time-integrated datasets set
strict limits on the fraction of the cosmic high-energy neutrinos that
can originate in these observed sources.

Possible alternative source populations include star-forming galaxies,
starburst galaxies, galaxy groups and clusters, supernovae, and
standard and low-luminosity gamma-ray bursts (see
\citealt{muraseorigin} for a review). Of these source possibilities,
the transient and highly-variable source populations will likely
require time-sensitive searches for identification.  Hadronic models
foresee that neutrinos and \grays\ are co-generated through the
production and subsequent decay of mesons, mainly pions.  \grays\ then
result from the decay of neutral pions, while the decay of charged
pions produces neutrinos. Additional processes in dense astrophysical
regions can then degrade the energy of individual \grays\ to lower
energies while leaving the neutrino energy spectrum almost unaffected,
resulting in correlated emission of higher-energy neutrinos and
lower-energy \grays.

The present paper is organized as follows: Details of the datasets are
provided in Sec.~\ref{sec:data}. Our statistical approach and signal
injection studies are discussed in Sec.~\ref{sec:meth}. Unscrambled
results and interpretation are presented in Sec.~\ref{sec:results},
and our conclusions in Sec.~\ref{sec:conc}.


\section{Datasets}
\label{sec:data}

The \fermi\ LAT dataset is highly complementary for cross-reference
with high-energy neutrino datasets. The LAT offers a 1.4~steradian
field of view, provides all sky coverage every three hours on average,
and exhibits good sensitivity over the ${\rm 100\,MeV} \simlt \egam
\simlt {\rm 300\,GeV}$ energy band.

This analysis was performed using publicly available \fermi\ LAT
data. The relevant \fermi\ data were the Pass~8 photon reconstructions
available from the LAT FTP server\footnote{LAT data located at
  \url{ftp://legacy.gsfc.nasa.gov/fermi/data/lat/weekly/photon/}}.
These photon events were filtered using the Fermi Science Tools,
keeping only photons with a zenith angle smaller than 90\arcdeg,
energies between 100~MeV and 300~GeV, detected during good time
intervals (GTI) as provided in the LAT satellite
files\footnote{\fermi\ satellite files located at
  \url{ftp://legacy.gsfc.nasa.gov/fermi/data/lat/weekly/spacecraft/}}.

The point spread function (PSF) of the LAT is given by a so-called
double King function \citep{kingfunc} with the parameters depending on
the photon energy, conversion type, and incident angle with respect to
the LAT boresight \citep{fermipsf}. At energies in the hundreds of
MeV, the angular uncertainty can be several degrees, especially for
off-axis photons. At $\egam > 1$\,GeV the average uncertainty drops
below 1\arcdeg, and at $\egam \simgt 100$\,GeV angular uncertainties
are better than 0.1\arcdeg.

The \antares\ data used spans from February 2007 to December
2015. Data from this 8.9 year interval are divided into track and
cascade events, all of which are upgoing. According to the selection
criteria defined in \citep{Albert:2017ohr}, during this period 7622
track and 180 cascade neutrino candidates were identified. The
\fermi\ mission has public data available starting from 4 August
2008. The \antares\ data is coincident with weeks 9 through 396 of the
\fermi\ data, with 6774 track-like events and 162 cascade-like events
falling within that 7.3 year window. For the \antares\ data, the
average PSFs for tracks and cascades are derrived from Monte-Carlo
simulation, and then interpolated. For track and cascacde events, the
90\% containment radii for the PSFs are 1\fdg 5 and
10\arcdeg\ respectively.

\begin{figure}
\includegraphics[width=\columnwidth]{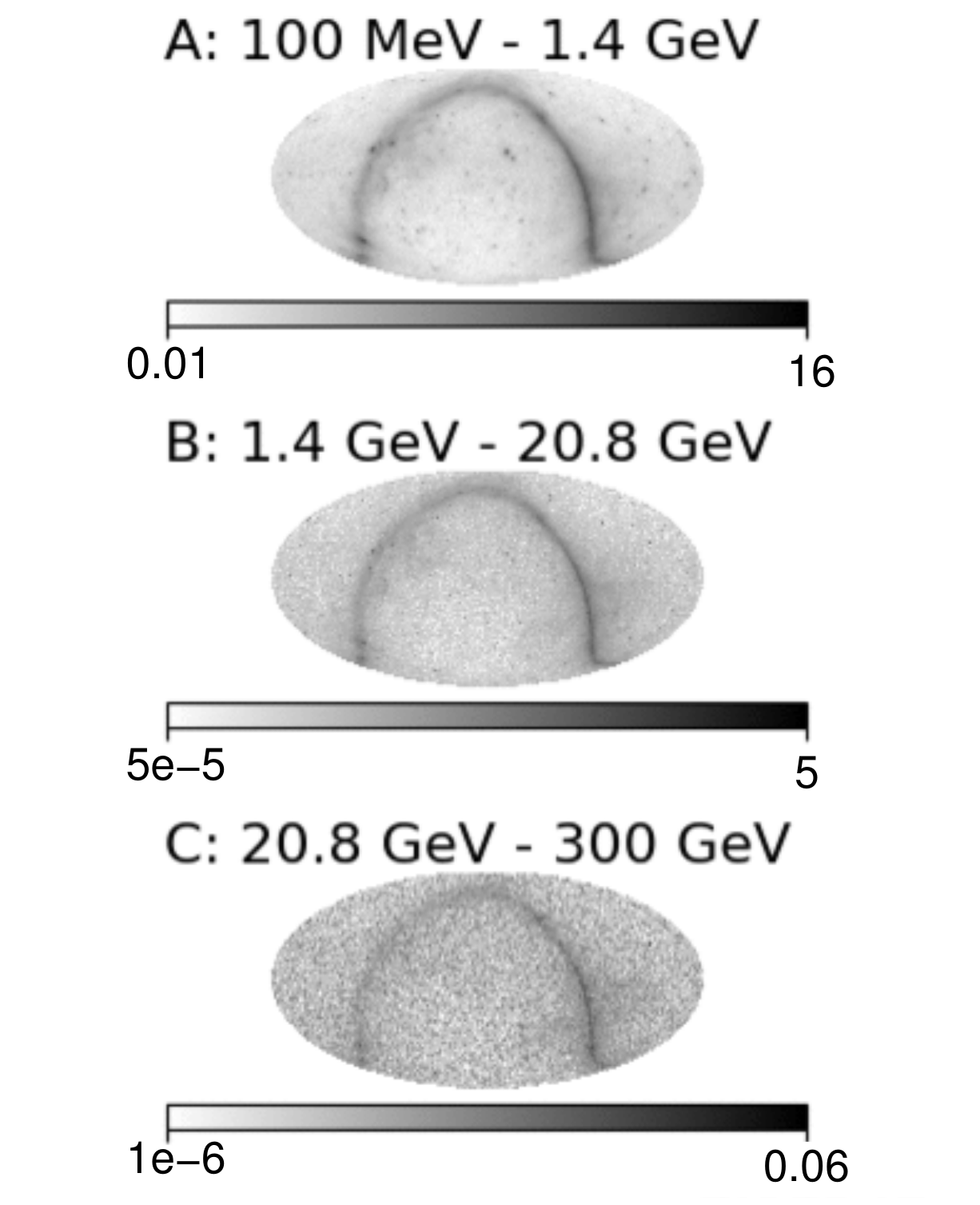}
\caption{Background maps of the \fermi\ LAT \gray\ sky. \fermi\ data
  are split into three logarithmically-uniform bins in energy and
  divided by the mission-averaged exposure map for that energy
  range. Grayscale intensity encodes the resulting mission-averaged
  photon flux over each band in units of photons per 200 seconds
  $m^{-2}$ ${\rm deg}^{-2}$.}
\label{fig:phbkg}
\end{figure}

A {\tt healpix} \citep{healpix} map of resolution 8 (NSide=256, mean
spacing of 0\fdg 23) was constructed using the entire \fermi\ data set
(weeks 9 to 495 at the time of creation) with aforementioned photon
selection criteria. Using the {\tt HEASoft} software\footnote{{\tt
    HEASoft} website:
  \url{https://heasarc.gsfc.nasa.gov/docs/software/lheasoft/}}, events
were binned into three logarithmically uniform energy bins. Each
energy bin was then further binned into a {\tt healpix} map, with the
live time calculated via a Monte Carlo simulation. Dividing the counts
map by the live time map produced the \fermi\ exposure
map. Zero-valued (low-exposure) pixels were replaced by the average of
the nearest neighbor pixels. Our three resulting all-sky \fermi\ maps
are shown in Fig.~\ref{fig:phbkg}. Due to the additional
reconstruction uncertainty in the \fermi\ PSF for high-inclination
events (inclination angle greater than 60\arcdeg), three additional
maps for analysis of these events were generated by further averaging
all pixels with their nearest neighbors.



\section{Methods}
\label{sec:meth}

\subsection{Significance Calculation}

\begin{figure}
\includegraphics[width=\columnwidth]{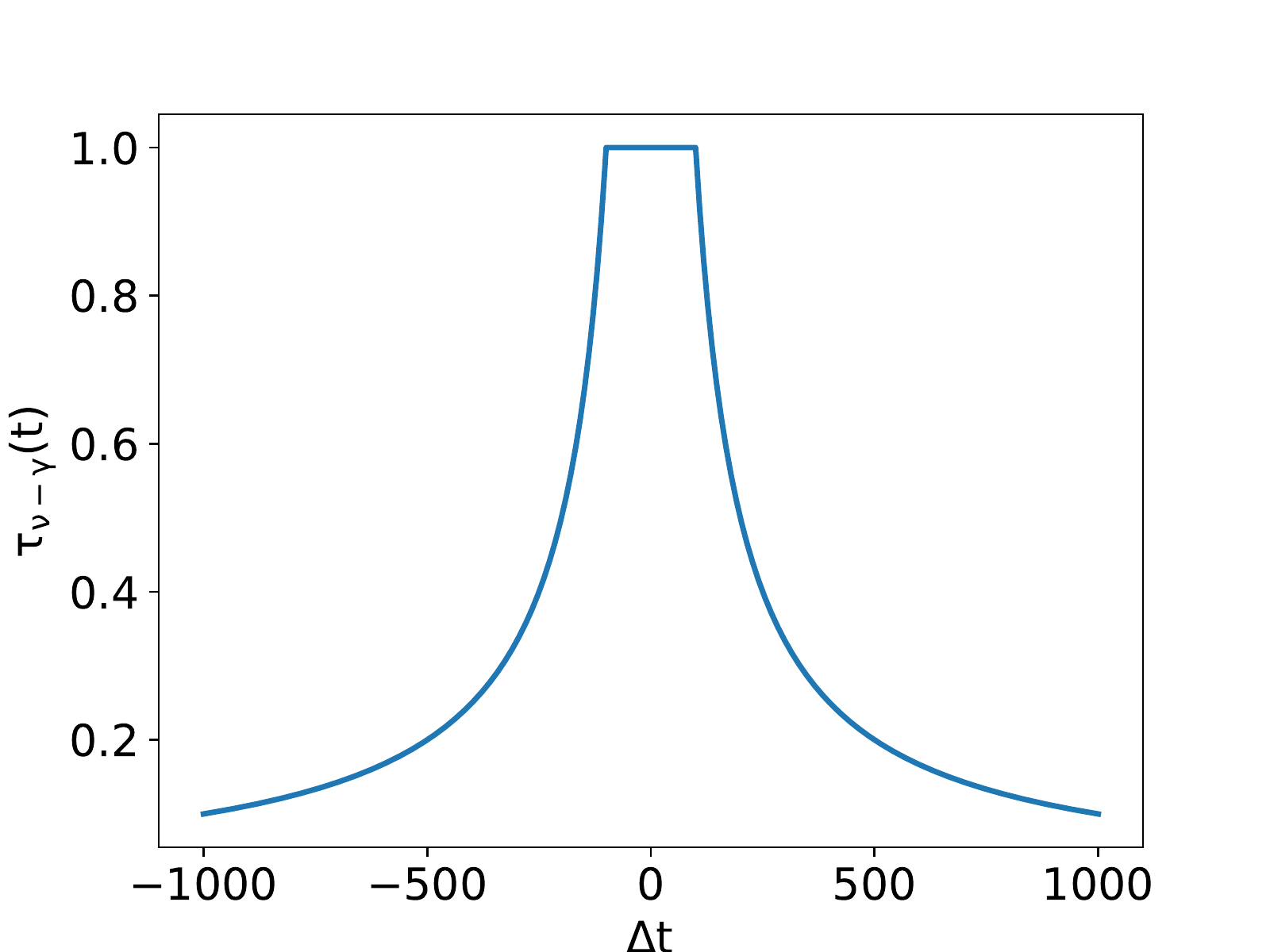}
\caption{Temporal weighting function $\tau(\Delta t)$ used in the
  analyses. For $|\Delta t|<$100~s, the function is flat and equal to
  1. For ${\rm 100\,s} < |\Delta t| < {\rm 1000\,s}$, the function
  scales as 1/$\Delta t$.}
\label{fig:tfunc}
\end{figure}

Our analysis follows as an extension to the methods presented in
\citet{turleyfermi18}. Different from previous work, our analysis
allows for coincidences with both multiple photons and multiple
neutrinos. Our analysis also covers both the track and cascade events
detected by \antares. For track-like events, we use an angular
acceptance window of 5\arcdeg, while for cascade-like events, we use
a 10\arcdeg\ window. For both event types, the temporal acceptance
window is $\pm$1000~s. Neutrino multiplets are constrained to have
each neutrino within both the angular and temporal separation of each
other neutrino. Photons must fall within the angular and temporal
window as measured from the average neutrino position and time. For
each coincidence, a pseudo-log-likelihood test statistic, $\lambda$,
is calculated as follows:
\begin{equation}
  \lambda= 2 \ln \frac{P_{\nu \gamma }(\vec{x}) \, n_\nu!\, n_\gamma! \,
    \Pi_{\nu,\gamma}\, \tau(\Delta t_i)}
   {\Pi_{\gamma}\, B_{\gamma,i}(\vec{x})} \\ +
    \sum_{\nu} \ln \frac{1 - p_{c,i}}{p_{c,i}},
\label{eq:lambda}
\end{equation}
%
%
%
where $P_{\nu \gamma}$ is the product of the point spread functions
(PSF) of each LAT photon and each \antares\ neutrino at the best
position, $\vec{x}$, with each PSF normalized to have units of
probability per square degree. The LAT PSF for each photon
additionally depends on the photon energy, inclination angle, and
conversion type. In general, the closer the PSF centers are, the
larger the resulting $\lambda$ value. The $n_\nu$ and $n_\gamma$ terms
are respectively the number of neutrinos and \grays\ in the
coincidence. The $\Pi_{\nu,\gamma}\, \tau(\Delta t_i)$ term is the
product of the temporal weighting function (Fig.~\ref{fig:tfunc})
evaluated for each neutrino and \gray\ in the coincidence.
For particles within 100\,s of the average arrival time, this function
is identically one, while it scales as $1/\Delta t$ for times between
100~s and 1000~s. This allows the search to address the possibility of
longer-timescale associations (as might result from low-luminosity
GRBs) while maintaining a preference for shorter-timescale
associations, if and when they are also present.

The $\Pi_\gamma\, B_{\gamma,i}(\vec{x})$ term is the product of LAT
\gray\ backgrounds for each photon at the coincidence location, taken
from the background maps shown in Fig~\ref{fig:phbkg}. Together with
the factorial terms, this acts like a Poisson probability of observing
$n_\gamma$ photons from background. The $p_c$ factor, similar to the
IceCube signalness \citep{icalertsys2017}, is an energy proxy
calculated by the ANTARES collaboration. The $p_c$ for a neutrino
event is computed on an event-by-event basis using the normalised
anti-cumulative distribution of the number of hits from the full
ANTARES 2012-2017 neutrino dataset. This probability represents the
fraction of ANTARES events with a number of hits larger than that
observed for the event: the larger the number of hits, the smaller the
$p_c$ value.  Overall, larger values of the $\lambda$ statistic
suggest a greater likelihood of a physically associated multiplet from
a cosmic source, rather than a coincidence of uncorrelated events.

The best fit position $\vec{x}$ is numerically calculated as the
location of maximum PSF overlap. The photon multiplicity of each
coincidence is calculated iteratively: Beginning with a coincidence
including all photons passing the temporal and proximity cuts, the
photon with the lowest PSF density at the best-fit position is removed
and a new $\lambda$, for the new best-fit position, is
calculated. This process is repeated until one photon is left
($n_\gamma$ iterations), with the iteration yielding the maximum
$\lambda$ selected as the coincidence multiplicity.

This analysis presents two ways to identify a potential signal. First,
with $\lambda$ unbounded, the null distribution provides threshold
values which can be used to identify individually-significant
coincidences and calculate their estimated false alarm rates. In this
work, we use two such thresholds, $\lambda_\mathrm{D}$ and
$\lambda_\mathrm{C}$, corresponding to false alarm rates of one per
decade and one per century, respectively.  Second, the presence of a
subthreshold population of \nugamma\ emitting sources can be
identified by a difference in the cumulative distributions of
$\lambda$ values between the observed and scrambled (null)
populations. By design, true coincidences will be biased to higher
$\lambda$ values, and a population containing a sufficient number of
signal events can be distinguished from the null distribution via an
Anderson-Darling $k$-sample test \citep{adtestpaper}.


\subsection{Background Generation}
\label{sub:search}

We generate a set of 10,000 Monte Carlo scrambled versions of each of
our datasets in order to characterize their null distributions and
define analysis thresholds, prior to performing any study of the
unscrambled datasets. Our scrambling procedure begins by first
converting the coordinates of each neutrino to detector coordinates.
The arrival time and azimuthal angle of each original neutrino $\nu_i$
are then exchanged with another randomly selected neutrino $\nu_j$.
Each neutrino retains its original elevation. Finally, the coordinates
are converted back to the equatorial system. This approach is similar
to the method used in our previous work \citep{turleyblazar}, with the
primary difference being the use of detector coordinates for the
scrambling procedure. \fermi\ LAT photons are not scrambled as the LAT
data contains known sources and extensive (complex)
structure. Coincidence analysis is carried out for each scrambled
dataset and $\lambda$ values are calculated for the resulting
\nugamma\ coincidences via Eq.~\ref{eq:lambda}.  Thresholds from this
analysis for false alarm rates of 1 per decade ($\lambda_{\rm D}$) and
1 per century ($\lambda_{\rm C}$) are presented in Table~\ref{tab:res}.


\begin{deluxetable*}{lrrrrrrrr}
  \tablecolumns{9}
  \tablecaption{Coincidence search results\label{tab:res}}

  \tablehead{%
    \colhead{~} &
    \colhead{~} &
    \multicolumn{4}{c}{Thresholds} &
    \multicolumn{3}{c}{Observed Values} \\
    \cline{7-9}
    \colhead{Dataset} &
    \colhead{$\langle n_{\nu+\gamma} \rangle$} &
    \colhead{$\lambda_{\rm D}$} &
    \colhead{$\lambda_{\rm C}$} &
    \colhead{$n_{\rm inj,1\%}$} &
    \colhead{$n_{\rm inj,0.1\%}$} &
    \colhead{$n_{\nu+\gamma}$} &
    \colhead{$\lambda_{\rm max}$} &
    \colhead{$p_{\rm A-D}$}} 
  
  \startdata 
  Tracks, 100~s    & $2716 \pm 36$   & 18.5 &   25.4 & 205 & 260 & 2734 & 18.94 & 39\% \\
  \phm{Tracks,} 1000~s &       ''    &  ''  &   ''   & 220 & 285 &  ''  &   ''  & ''   \\
  Cascades         & $83.6 \pm 5.8$  &  8.1 &   14.6 & -   & -   &   80 & 2.7   & 60\% \\
  Track Multiplets & $0.48 \pm 0.69$ & -    & $-$9.3 & -   & -   &    0 &   -   &  -   \\
  \enddata

  \tablecomments{$\langle n_{\nu+\gamma} \rangle$ is the expected
    number of neutrinos observed in coincidence with one or more
    \grays, as derived from 10,000 Monte Carlo scrambled realizations
    of each dataset. \lamd\ and \lamc\ are the thresholds above which
    a coincidence is observed only once per simulated decade or
    century, respectively. $n_{\rm inj,1\%}$ and $n_{\rm inj,0.1\%}$
    are the number of injected signal events required in simulations
    to give Anderson-Darling test \citep{adtestpaper} $p$-values of
    $p<1\%$ and $p<0.1\%$, respectively, by comparison to the null
    distributions for each dataset. $n_{\nu+\gamma}$ is the number of
    neutrinos observed in coincidence with one or more \grays\ in
    unscrambled data, $\lambda_{\rm max}$ is the maximum observed
    $\lambda$ for each dataset, and $p_{\rm A-D}$ is the
    Anderson-Darling test $p$-value from comparison of the observed
    $\lambda$ distribution to the associated null distribution. Cells
    with a `-' could not be calculated, for reasons detailed in the
    main text.}

\end{deluxetable*}


In contrast to previous work \citep{turleyfermi18}, due to the
sensitivity to multi-neutrino events and the use of both track and
cascade events, we split the analysis into three separate parts. The
first part is to detect all coincidences with single-neutrino
track-like events.  The second looks for coincidences with
multi-neutrino track like events.  The third and final part is a
search for coincidences with all single-neutrino cascade-like
events. Multi-neutrino cascades are not considered, as there are no
cascade-like events within the temporal acceptance window of each
other.


\subsection{Signal Injection}
\label{sub:injection}
To estimate the sensitivity of our analysis to subthreshold
populations of cosmic \nugamma\ emitting sources, we generate a
population of signal-like events. These events are injected into the
scrambled datasets so that the injected distributions can be compared
to the null distribution.

We determine the multiplicity of a generated signal event following
the methods used in \citet{turleyfermi18}. This method assumes a
population of sources emitting one neutrino, with associated photon
fluence distributed according to $N(S\ge S_0) \propto S_{0}^{-3/2}$.
In this formulation, $N(S\ge S_0)$ is the number of events observed
with a fluence greater than the threshold fluence $S_0$.  Setting this
minimum to 0.001 photons, we can invert this relationship and generate
the expectation value for the multiplicity of an arbitrary event in
terms of a uniform random variable $u$ as $\ngexp = S_{0}\,
u^{-2/3}$. The distribution of $n_\gamma$ is then calculated by
drawing randomly from a Poisson distribution with the expectation
value \ngexp. Excluding events with zero photons, this yields the
following $n_\gamma$ distribution: 93.8\% singlet, 4.5\% doublet,
0.9\% triplet, and 0.38\%, 0.19\%, 0.095\%, 0.0567\%, 0.0365\%,
0.0244\%, and 0.0174\% for multiplicities four through ten.

A signal event of photon multiplicity $n_\gamma$ is then generated by
choosing a random right ascension and drawing a random declination
from the list of all \antares\ events. These coordinates serve as a
sky position for the coincidence. The PSFs for $n_\gamma$ LAT photons
and $n_\nu$ neutrinos are then centered on this point, and placed
randomly according to their respective PSFs. All photons are chosen to
have the same inclination angle, which is drawn from the full set of
inclination angles within the \fermi\ dataset. A conversion type for
each photon is similarly drawn from the \fermi\ dataset.  Photon
energies are drawn from a power law with a photon index
$\Gamma=2$. Using the photon background maps, the number of
unassociated photons expected to arrive within the temporal and
spatial windows for that section of sky is calculated.  From this
Poisson probability, $n_b$ photons are randomly placed uniformly
within the spatial window. Energy and conversion type for the
background photons are chosen in the same manner as for the signal
photons. All background photons are given the same inclination angle
as the signal photons. Each particle is also given an arrival time
randomly selected from a uniform distribution. Using this information,
a $\lambda$ value is calculated following the methods of
Sec~\ref{sec:meth}. Due to the iterative rejection of one or more
low-significance \grays, events can end up with some of the injected
photons excluded.

Because the varied physical models predicting \nugamma\ coincidences
have different characteristic timescales, we generate two sets of
signal events for each of the three null distributions. One set draws
the timestamps from a uniform distribution 100~s wide, while the other
draws from a uniform distribution 1000~s wide.

\begin{figure*}
\includegraphics[width=\columnwidth]{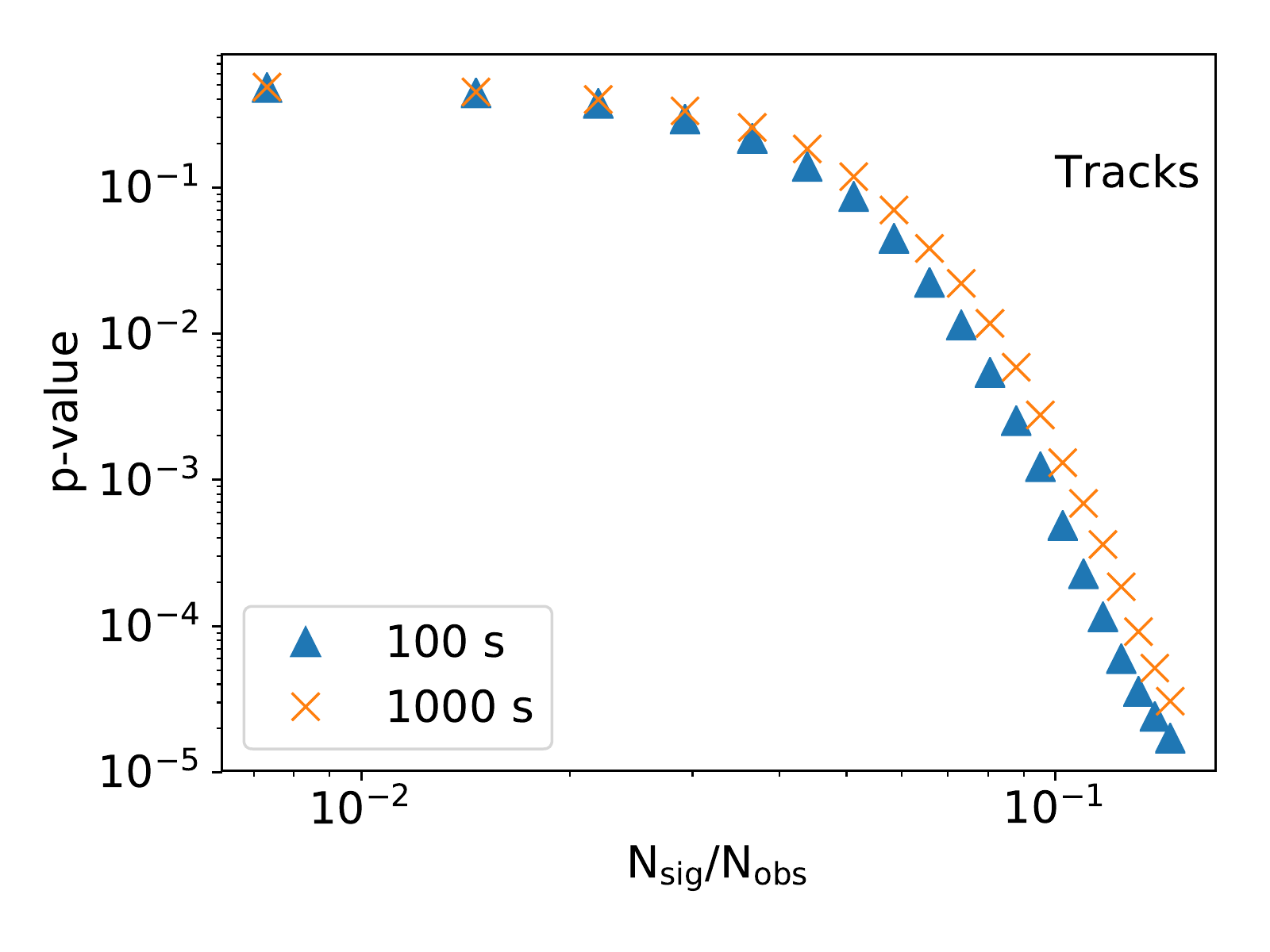}
\includegraphics[width=\columnwidth]{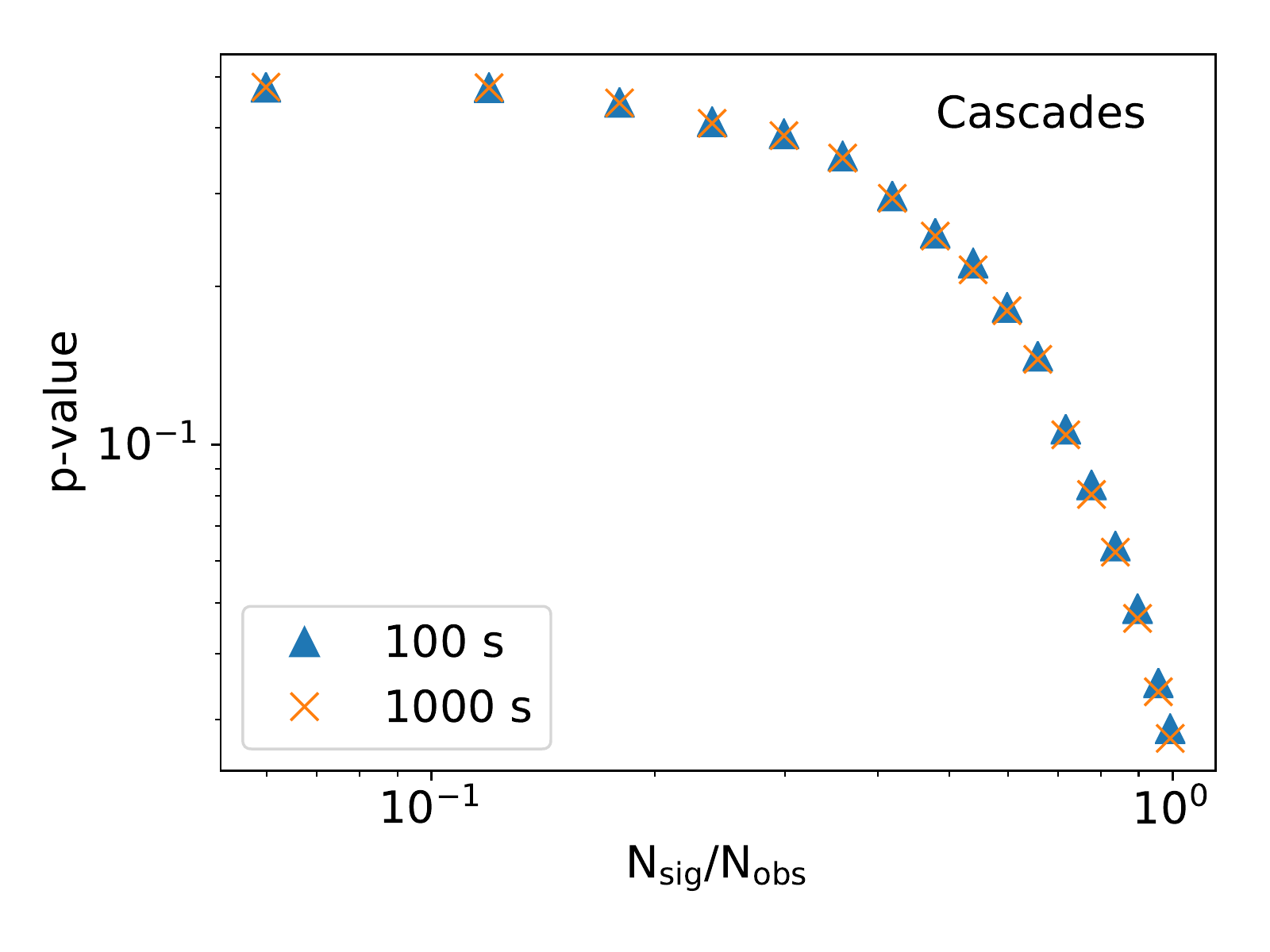}
\caption{Anderson-Darling two-sample $p$-value versus fraction of
  coincidences that result from signal events, $N_{\rm sig}/N_{\rm
    obs}$. Results from both signal populations are shown. }
\label{fig:ad}
\end{figure*}

To calculate the sensitivity of our analysis, we inject an increasing
number of signal events \ninj\ and plot the median resulting
Anderson-Darling $p$-value \citep{adtestpaper} against \ninj$/n_{\rm
  obs}$ for the track and cascade data, as shown in Fig.~\ref{fig:ad}.

For the tracks, this provides an estimate of the threshold value of
\ninj\ that is needed to yield a statistically significant deviation
from the null distribution (see columns $n_{\rm inj,1\%}$ and $n_{\rm
  inj,0.1\%}$ in Table~\ref{tab:res}). For the cascades, the size of
each individual scramble is small enough that replacing 100\% of the
dataset with signal events yields a $p$-value of 2.8\% on average,
making it very unlikely that this sample would yield a high-confidence
demonstration of an underlying \nugamma\ source population. At 90\%
confidence, our analysis is sensitive to $>$130 source-like
\nugamma\ coincidences in the 100~s track data, $>$145 in the 1000~s
track data, and $>$60 in the 100~s and 1000~s cascade data.  Relevant
statistics from these analyses are provided in Table~\ref{tab:res}.

In previous work, \citet{turleyfermi18} found that scrambled neutrinos
coincident with LAT-detected GRBs, in particular GRB\,090902B
\citep{latgrb090902}, yielded $\lambda$ values well above the
\lamc\ threshold. To quantify our analysis sensitivity to
GRB + neutrino coincidences, we carried out a Monte-Carlo simulation
for each LAT-detected GRB\footnote{LAT GRB catalog:
  \url{https://fermi.gsfc.nasa.gov/ssc/observations/types/grbs/lat_grbs/}}
that occurred within our data collection period. Neutrinos were
injected following our signal injection procedures, with the GRB
position and trigger time as reference, and with a 1000-second
box-window temporal distribution for neutrino arrival times.  For each
LAT GRB, we carried out 10,000 such neutrino signal injections and
calculated the $\lambda$ value for the resulting association in each
instance.

The maximum $\lambda$ generated through this search was $\lambda =
3524.5$, resulting from a 368-photon coincidence with GRB\,130427A
\citep{latgrb130427}. Of the 128 individual bursts in this simulation,
58 have median $\lambda$ values from these neutrino injection trials
of $\lambda_{\rm med}>\lamc$, and a further five bursts have $\lamc >
\lambda_{\rm med} > \lamd$.


\section{Results}
\label{sec:results}

\begin{figure*}
\includegraphics[width=\columnwidth]{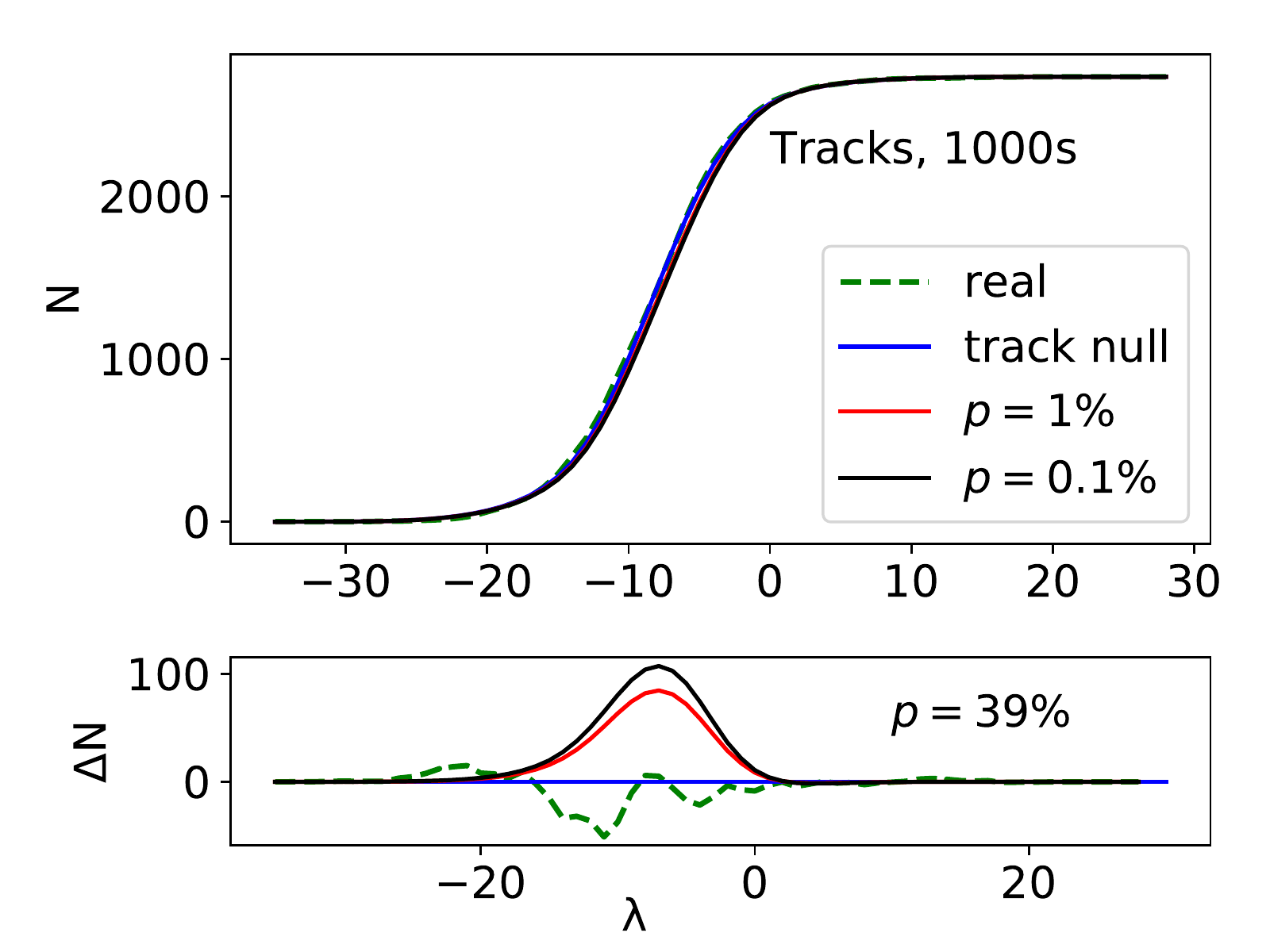}
\includegraphics[width=\columnwidth]{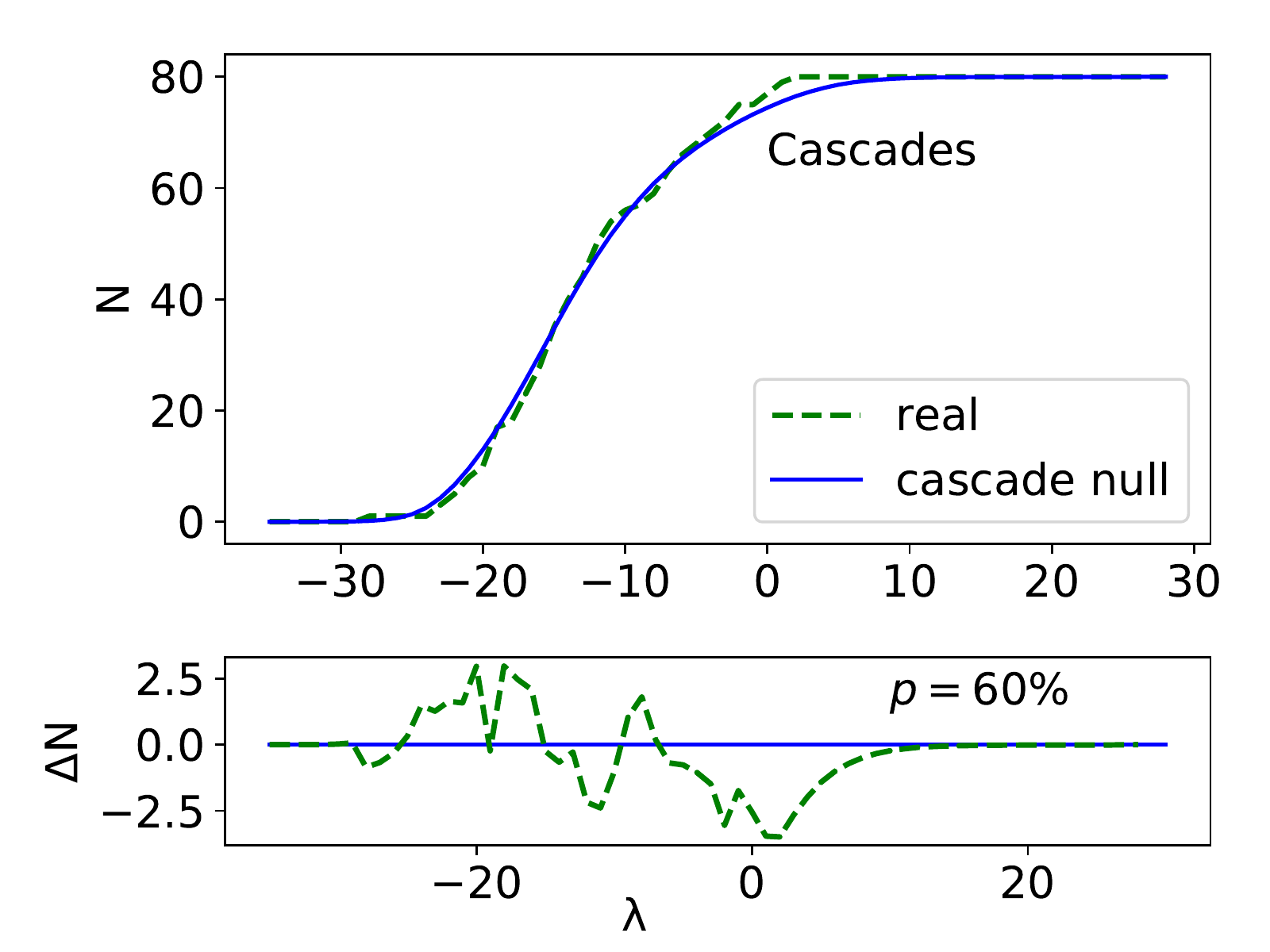}
\caption{Cumulative and residual histograms of the $\lambda$
  distributions for the track (left, $n_{\nu+\gamma}=2734$) and
  cascade (right, $n_{\nu+\gamma}=80$) data. The unscrambled data
  (green dashed line) and the null distribution (blue line) are shown
  for both tracks and cascades. Signal injections, generated using a
  1000~s temporal window and yielding $p=1\%$ (red line) and $p=0.1\%$
  (black line) are calculated for the track data only, as even 100\%
  signal injection does not allow strong discrimination of signal and
  null distributions for the cascade data. Signal injection curves for
  the 100~s temporal window display as identical on this plot. Upper
  panels show cumulative histograms, while lower panels show residuals
  against the null distribution (plotted as null minus
  alternative). Anderson-Darling test $p$-values from comparison of
  the unscrambled and null distributions are $p=39\%$ for the track
  sample and $p=60\%$ for the cascade sample.}
\label{fig:res}
\end{figure*}

Applying our analysis to the two unscrambled neutrino datasets yields
the results summarized in Table~\ref{tab:res}. Fig.~\ref{fig:res}
shows the $\lambda$ distributions for the unscrambled data for the
track and cascade data, along with the null distributions, and
distributions for signal injections (where possible) yielding
$p$-values of 1\% and 0.1\%, respectively.
All distributions are normalized to the number of coincidences in the
unscrambled distribution. Note that due to the small size of the
cascade coincidence sample, it is not possible to inject enough signal
events into a random scramble to differentiate from other random
scrambles at better than $p$=2.8\% (97.2\% confidence).

Two coincidences above the \lamd\ threshold were observed in the track
data. From Poisson statistics, two or more such coincidences would be
observed 16.6\% of the time given the 7.3 year span of the
data. Details of these two coincidences are presented in
Table~\ref{tab:high}. No $\lambda$ values above the \lamd\ threshold
were observed in the cascade data. The subthreshold population search
demonstrated that both unscrambled distributions were consistent with
background, with test statistics of 39\% for the tracks, and 60\% for
the cascades. Results from the track multiplet analysis are not shown
as there were, on average, only 0.48 such coincidences per scramble,
and none in the unscrambled analysis.


\begin{deluxetable*}{lrrrrrrrr}
  \tablecolumns{8}
  \tablecaption{High-$\lambda$ events\label{tab:high}}

  \tablehead{
    \colhead{Date} &
    \colhead{Time (UTC)} &
    \colhead{MJD} &
    \colhead{$\Delta$t (s)} &
    \colhead{Position (J2000)} &
    \colhead{$r_{1\sigma}$} &
    \colhead{$N_{\rm ph} $} &
    \colhead{$\lambda$} &
    \colhead{FAR (\peryear)}}  

  \startdata
  2012 Nov 21 & 20:19:52 & 56252.8471 & 307 & $248\fdg 00, -7\fdg 70$ &
                2\arcmin & 1 & 18.9 & 0.09 \\
  2014 Aug 05 & 11:13:33 & 56874.4677 & 750 & $279\fdg 68, -5\fdg 05$ &
                3\arcmin & 2 & 18.8 & 0.09 \\
  \enddata

  \tablecomments{Date, Time, and MJD show the central time of the
    coincidence, while $\Delta$t measures the separation between the
    earliest and latest particles in the coincidence in
    seconds. Position gives the RA and Dec (in degrees) of the best
    fit position, while $r_{1\sigma}$ gives the approximate 1$\sigma$
    error on the angular uncertainty in arcminutes (39\% containment,
    assuming a Gaussian form). $N_{\rm ph}$ is the number of photons in
    the coincidence. The false alarm rate (FAR) is calculated as the
    number of events of that $\lambda$ or higher expected per year.}
  
\end{deluxetable*}

\begin{figure}
\includegraphics[width=\columnwidth]{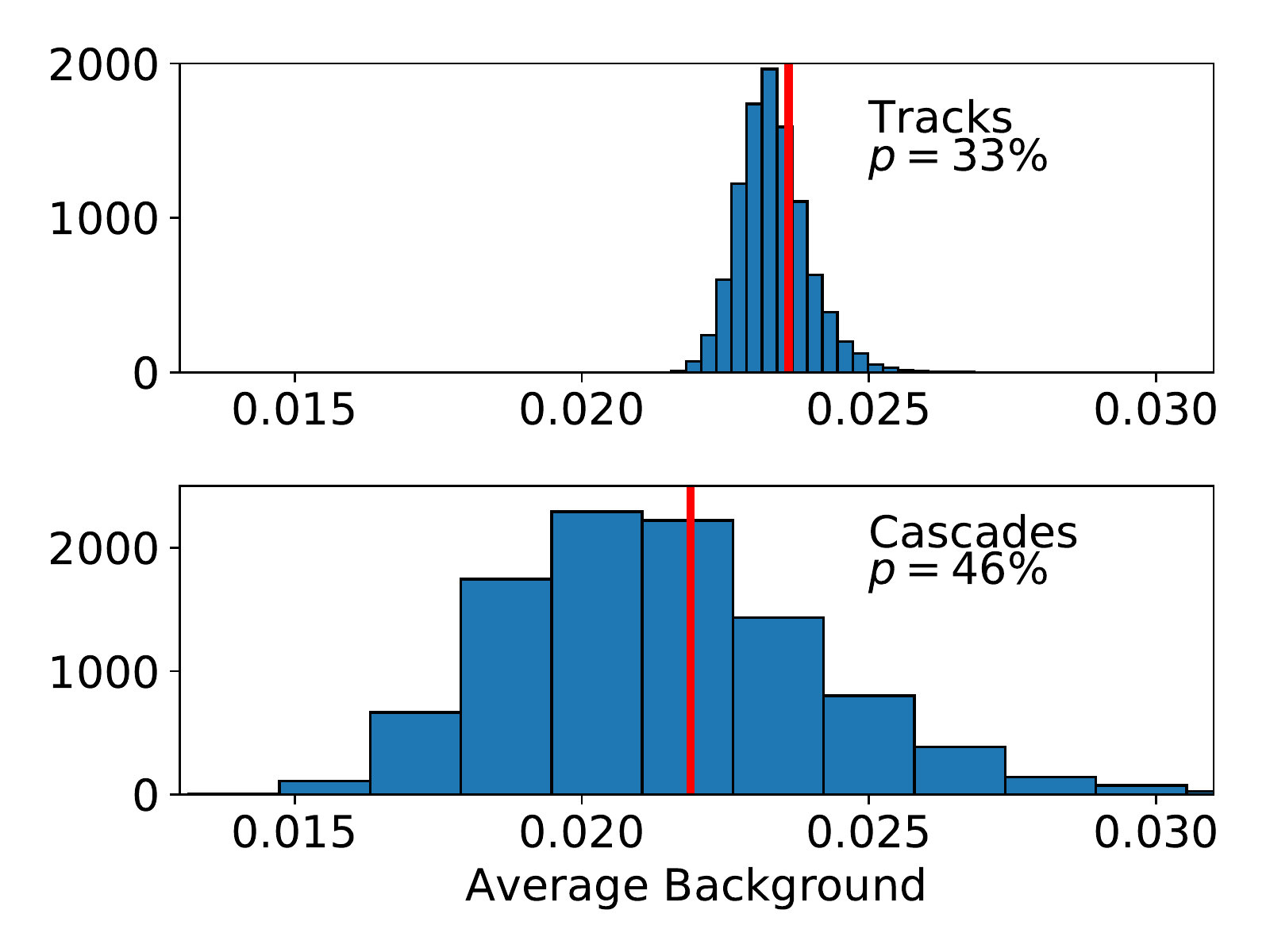}
\caption{Average \fermi\ \gray\ background rates at the positions
  of track (upper panel) and cascade (lower panel) neutrinos. In
  each panel, the histogram shows the distribution obtained from
  10,000 Monte-Carlo scrambled datasets, while the red line marks the
  observed background rate for unscrambled data. Background rates are
  expressed in units of photons per square meter per square degree per
  200s. Observed average backgrounds are consistent with background
  for both datasets.}
\label{fig:bkgtest}
\end{figure}

\citet{turleyfermi18} also tested for correlation between neutrino and
\fermi\ LAT photon sky positions without any temporal
correlation. Repeating this analysis using the \antares\ data, we
first construct a single \fermi\ background map covering the full
energy range. We then measure the background value at the location of
every neutrino in the track and cascade data to compute an average
photon background for each neutrino map. Carrying this out on the
scrambled neutrino datasets yields an average background of $(2.33 \pm
0.06)\times 10^{-2} $ photons deg$^{-2}$ m$^{-2}$ per 200~s for the
track data, and $(2.16 \pm 0.36)\times 10^{-2} $ photons deg$^{-2}$
m$^{-2}$ per 200~s for the cascade data. The observed backgrounds (in
the same units) from the unscrambled data are $2.36\times 10^{-2}$
(+0.44 $\sigma$; $p$ = 33\%) for the track data, and $2.19\times
10^{-2}$ (+0.09 $\sigma$; $p$ = 46\%) for the cascade data. Both
results are consistent with background (Fig.~\ref{fig:bkgtest}.)
The dispersion in the cascade background from scrambled datatsets is
far larger than that for the tracks because of the much-reduced sample
size (180 cascade events compared to 7622 track events); however, the
two average backgrounds are consistent, as the mean of the track
background is 0.47$\sigma$ larger than the mean of the cascade
background, as measured using the standard deviation of the cascade
background distribution.  Recalling the IC59 Northern ($p$=28.1\%),
IC59 Southern ($p$=4.7\%), and IC40 ($p$=58.3\%) results from
\citet{turleyfermi18}, we can calculate a unified $p$-value of 19.7\%
from these values using Fisher's method \citep{fisher48}.

\vspace*{\baselineskip}


\section{Conclusions}
\label{sec:conc}

We have carried out a search for \nugamma\ transients using publicly
available \fermi\ LAT \gray\ data and \antares\ neutrino data.  Our
analysis used archival data from both observatories over the period
August~2008 to December~2015. As with previous work
\citep{turleyfermi18}, our analysis was designed to be capable of
identifying either individual high-significance \nugamma\ transients
or a population of individually subthreshold events, via statistical
comparison to uncorrelated (scrambled) datasets.

Our Monte Carlo simulations demonstrate a sensitivity to
single-neutrino events of sufficient \gray\ multiplicity, as
demonstrated by signal injection against multiple bright LAT-detected
\gray\ bursts. Signal injection against scrambled datasets established
our sensitivity to subthreshold populations of transient
\nugamma\ sources at the $>$7\% level ($>$200 coincidences) for
tracks; however, due to the small sample size, we were not able to
place meaningful limits on a subthreshold \nugamma\ source population
within the cascades data. Our limit of $>$200 coincidences in the full
dataset is equivalent to $>$27 LAT-associated cosmic neutrinos per
year in the \antares\ data. Since \ice\ estimates of the cosmic
neutrino flux and spectrum lead us to expect 6.8~cosmic
\antares\ neutrinos per year (Sec.~\ref{sec:intro}), our limit is
not physically constraining in this context.

Analysis of the observed (unscrambled) data reveals two
\nugamma\ coincidences above a nominal \lamd\ threshold (false alarm
rate ${\rm FAR} < 0.1$\,\peryear; Table~\ref{tab:high}). Due to the
7.3~year span of the data, we anticipate observing two or more
$\lambda > \lamd$ coincidences 16.6\% of the time ($p=16.6\%$). We
observe no statistically-significant deviation of the observed
$\lambda$ distributions from their associated null distributions, with
observed $p$-values of $p=39\%$ and $p=60\%$ for the track and cascade
events, respectively.

Independently, we performed the first test for correlation between
\antares\ neutrino positions and persistently bright portions of the
\fermi\ \gray\ sky. Our test found no significant excess in either
the tracks ($p=33\%$) or cascades ($p=46\%$). Combining these values
with previous results (28.1\% for IC59 north, 4.7\% for IC59 south,
58.3\% for IC40; \citealt{turleyfermi18}) by Fisher's method yields a
joint $p$-value of $p=19.7\%$.

While our results show no significant evidence of
\nugamma\ coincidences, we look forward to the results of future
searches using additional neutrino data. We also continue our work
with Astrophysical Multimessenger Observatory Network
\citep{amondesg,amonrt} partner facilities and the Gamma-ray
Coordinates Network \citep{gcn98} to generate low-latency
\nugamma\ alerts from \fermi\ LAT \gray\ and high-energy neutrino
data. Once these alerts are deployed, they will be distributed in real
time to AMON follow-up partners.



\acknowledgments

The authors thank David Thompson for helpful discussions. We
gratefully acknowledge support from Penn State's Office of the Senior
Vice President for Research, the Eberly College of Science, and the
Penn State Institute for Gravitation and the Cosmos. This work was
supported in part by the National Science Foundation under Grant
Number \mbox{PHY-1708146}. K.~M. is supported by the Alfred P. Sloan
Foundation and by the National Science Foundation under Grant Number
\mbox{PHY-1620777}.
The authors acknowledge the financial support of the funding agencies:
Centre National de la Recherche Scientifique (CNRS), Commissariat \`a
l'\'ener\-gie atomique et aux \'energies alternatives (CEA),
Commission Europ\'eenne (FEDER fund and Marie Curie Program),
Institut Universitaire de France (IUF), IdEx program and UnivEarthS
Labex program at Sorbonne Paris Cit\'e (ANR-10-LABX-0023 and
ANR-11-IDEX-0005-02), Labex OCEVU (ANR-11-LABX-0060) and the
A*MIDEX project (ANR-11-IDEX-0001-02),
R\'egion \^Ile-de-France (DIM-ACAV), R\'egion
Alsace (contrat CPER), R\'egion Provence-Alpes-C\^ote d'Azur,
D\'e\-par\-tement du Var and Ville de La
Seyne-sur-Mer, France;
Bundesministerium f\"ur Bildung und Forschung
(BMBF), Germany; 
Istituto Nazionale di Fisica Nucleare (INFN), Italy;
Nederlandse organisatie voor Wetenschappelijk Onderzoek (NWO), the Netherlands;
Council of the President of the Russian Federation for young
scientists and leading scientific schools supporting grants, Russia;
Executive Unit for Financing Higher Education, Research, Development
and Innovation (UEFISCDI), Romania;
Mi\-nis\-te\-rio de Econom\'{\i}a y Competitividad (MINECO): Plan
Estatal de Investigaci\'{o}n (refs. FPA2015-65150-C3-1-P, -2-P and
-3-P, (MINECO/FEDER)), Severo Ochoa Centre of Excellence and Red
Consolider MultiDark (MINECO), and Prometeo and Grisol\'{i}a programs
(Generalitat Valenciana), Spain;
Ministry of Higher Education, Scientific Research and Professional
Training, Morocco.
We also acknowledge the technical support of Ifremer, AIM and Foselev Marine
for the sea operation and the CC-IN2P3 for the computing facilities.

\software{Astropy \citep{astropy}, Matplotlib \citep{matplotlib},
  HEASoft \citep{heasoft}, HEALPix \citep{healpix}, SciPy \citep{scipy}}


\bibliographystyle{aasjournal} \bibliography{fic}


\end{document}